\documentclass[twocolumn,superscriptaddress,aps,pra]{revtex4-1}
\usepackage{bm}
\usepackage{appendix}
\usepackage{graphicx}
\usepackage{amsmath}
\usepackage{amssymb}%
\usepackage{color}
\usepackage{lipsum}
\usepackage{multirow}
\usepackage[colorlinks=true,citecolor=blue,urlcolor=blue]{hyperref}
\usepackage{mathrsfs}
\usepackage{braket}
\usepackage{hhline}
\usepackage{ulem}
\usepackage{hyperref}
\usepackage{xurl}

\begin{document}
		
   \title{Non-Hermitian skin effect in one-dimensional interacting Bose gas}
	
	\author{Liang Mao}
	\affiliation{Institute for Advanced Study, Tsinghua University, Beijing, 100084, China}
	\author{Yajiang Hao}
	\affiliation{Department of Physics, University of Science and Technology Beijing, Beijing 100083, China}
	\author{Lei Pan}
	\email{panlei@mail.tsinghua.edu.cn}
	\affiliation{Institute for Advanced Study, Tsinghua University, Beijing, 100084, China}

\begin{abstract}
Non-Hermitian skin effect (NHSE) is a unique feature studied extensively in non-interacting non-Hermitian systems. In this work, we extend the NHSE originally discovered in non-interacting systems to interacting many-body systems by investigating an exactly solvable non-Hermitian model, i.e., the prototypical Lieb-Liniger Bose gas with imaginary vector potential. We show that this non-Hermitian many-body model can also be exactly solved through Bethe ansatz. By solving the Bethe ansatz equations accurately, the explicit eigenfunction is obtained, and the model's density profiles and momentum distributions are calculated to characterize the NHSE quantitatively. We find that the NHSE is gradually suppressed on the repulsive side but does not vanish as the repulsive interaction strength increases. 
On the attractive side, the NSHE for bound-state solutions is enhanced as interaction strength grows.
In contrast, for the scattering state the NHSE shows a non-monotonic behavior in the attractive side. Our work provides the first example of the NHSE in exactly solvable many-body systems, and we envision that it can be extended to other non-Hermitian many-body systems, especially to integrable models.	 
		
	\end{abstract}
	
	\maketitle
	
	
\section{Introduction}
Open quantum systems are ubiquitous in nature, which is an essential branch of modern physics and has penetrated into numerous areas, including atomic and molecule physics, nuclear physics, photonics, biophysics, mesoscopic physics, etc. 
Non-Hermitian Hamiltonians can describe an open quantum system effectively. The first example was introduced by George Gamow, who derived a non-Hermitian model to describe the alpha decay of heavy nuclei \cite{Gamow}. With the discovery of $\mathcal{PT}$-symmetry \cite{Bender} in non-Hermitian Hamiltonian and associated experimental observation of $\mathcal{PT}$-symmetry breaking, non-Hermitian physics has attracted intense attention.

In recent years, non-Hermitian physics has roused revived theoretical interest due to tremendous advances in experimental technology for controlling dissipation \cite{NH_Review1,NH_Review2}. 
Thanks to the great advantage in manipulating atom-atom interaction and light-matter coupling, ultracold atom experiments provide an unprecedented opportunity to investigate interacting non-Hermitian systems \cite{Exp1,Exp2,Exp3,Exp4,Exp5,Exp6,Exp7,Exp8,Exp9,Exp10}.	
Recent theoretical studies in non-Hermitian many-body systems have revealed that the interplay between interaction and non-Hermiticity can alter physical properties and give rise to intriguing phenomena absent in Hermitian many-body systems   \cite{NHMB1,NHMB2,NHMB3,NHMB4,NHMB405,NHMB5,NHMB6,NHMB7,NHMB8,NHMB9,NHMB11,NHMB12,NHMB125,NHMB126,NHMB127,NHMB13,NHMB14,NHMB15,NHMB16,NHMB17,NHMB18,NHMB19,NHMB20,NHMB21,NHMB22,NHMB23,NHMB24,NHMB25,NHMB26,NHMB27,NHMB28,NHMB29,NHMB30,NHMB301,NHMB302,NHMB303,NHMB31,NHMB32,NHMB33,NHMB34,NHMB35,NHMB355,NHMB356,NHMB36,NHMB37,NHMB38,NHMB39,NHMB40,NHMB41,NHMB42,NHMB43,NHMB44,NHMB45,NHMB46,NHMB47,NHMB48} such as non-Hermitian superfluidity  \cite{NHMB4,NHMB405,NHMB5}, non-Hermitian quantum magnetism \cite{NHMB24,NHMB33}, and non-Hermitian many-body localization   \cite{NHMB35,NHMB355,NHMB356}, etc.
 

A unique feature of non-Hermitian systems in open boundary conditions is the so-called non-Hermitian skin effect (NHSE)   \cite{Skin_Theory1,Skin_Theory2,Skin_Theory3} which is recognized by eigenfunctions accumulated at a boundary, akin to the charge distribution over the surface in a conductor.
More recently, non-Hermitian systems featuring the NHSE have been raising an increasing concern \cite{Skin_Theory1,Skin_Theory2,Skin_Theory3,Skin_Theory4,Skin_Theory5,Skin_Theory6,Skin_Theory7,Skin_Theory8,Skin_Theory9,Skin_Theory10,Skin_Theory11,Skin_Theory12,Skin_Theory125,Skin_Theory13,Skin_Theory14,Skin_Theory15,Skin_Theory16,Skin_Theory17,Skin_Theory18,Skin_Theory185,Skin_Theory19,Skin_Theory20,Skin_Theory21,Skin_Theory22,Skin_Theory23,Skin_Theory24,Skin_Theory245,Skin_Theory246} motivated by the experimental realization of NHSE \cite{Skin_Exp1,Skin_Exp2,Skin_Exp3,Skin_Exp4,Skin_Exp5,Skin_Exp6,Skin_Exp7,Skin_Exp8,Skin_Exp9}. Despite extensive investigations, the current studies of the NHSE discussed in the literature mainly focus on the single-particle level, such as non-Hermitian topological bands or non-Hermitian quasicrystals.
In contrast, the research on NHSE in interacting systems is in its infancy \cite{Skin_Theory23,Skin_Theory24,Skin_Theory25,Skin_Theory26}, which just involves few-body calculation \cite{Skin_Theory22}, exact diagonalization study \cite{Skin_Theory23,Skin_Theory24,Skin_Theory25}, perturbation theory \cite{Skin_Theory26}, and hard-core limit \cite{Skin_Theory27}. The NHSE in exactly solvable many-body systems has
not been systematically investigated so far.   

In this paper, we theoretically investigate NSHE in an exactly solvable model. Exactly solvable models play a significant role in stastistical physics and condensed matter physics, such as the verification of Bogoliubov theory in Lieb-Liniger model \cite{Lieb-Liniger} and Wilson's numerical renormalization group in exact solutions of Kondo model \cite{Kondo1}.
With the help of techniques developed in the integrable model literature, we can obtain exact solutions of eigen-energies and wavefunctions. The NHSE sensitively depends on boundary conditions, which is similar to integrability conditions in many-body systems. So we need to consider a non-Hermitian interacting system with open boundary condition (OBC), which manifests the NHSE and meanwhile guarantees the integrability condition. 
Based on this criterion, we employ the 1D interacting Bose gas (Lieb-Liniger model) under OBC with an additional imaginary potential corresponding to the nonreciprocal hopping in the 1D lattice, see (\ref{H}). This model has different physical properties as interaction strength varies, and can form bound states on attractive interaction side. We will investigate NSHE and its response to interaction in the whole interaction range.

The rest of this paper is organized as follows. In Sec. \ref{Sec. Model}, we introduce the exactly solvable non-Hermitian many-body model and give the exact solution obtained by Bethe ansatz, where the explicit eigenfunction and eigenvalue are derived.
In Sec. \ref{Sec. repulsive}, we discuss the NHSE in a repulsive interaction regime where both density profiles and momentum distributions are calculated to quantify degree of the NHSE. Then we explore the NHSE in attractive interaction, including the bound state and scattering state in Sec. \ref{Sec. attractive}.  
Finally, we summarize this paper in Sec. \ref{Summary}. For readers who are not familiar with NSHE, Appendix \ref{App1} provides an introductory example for NSHE in a tight-binding model.

\section{Model and solution}
\label{Sec. Model}
In this section, we introduce an exactly solvable non-Hermitian many-body model and then carry out the Bethe ansatz solution. We focus on a 1D system that consists of $N$ bosons with $\delta$-function interaction and subjects to an imaginary potential 
\begin{eqnarray} 
\hat{H}=\sum_{j=1}^{N} \Big[-i\frac{\partial}{\partial x_{j}}+i\phi(x_j)\Big]^2+2 c \sum_{j<l} \delta\left(x_{j}-x_{l}\right),\label{H}
\end{eqnarray}
where $c$ denotes the interaction strength and $i\phi(x)$ is the imaginary potential. Here we set $\hbar=2m=1$. This model is known as Lieb-Liniger model \cite{Lieb-Liniger} when $\phi(x)=0$ and has been realized in ultracold atomic gases \cite{Lieb_Exp1,Lieb_Exp2,Lieb_Exp3}. The interaction strength $c$ can be tuned by the confinement
induced resonance \cite{CIR} or Feshbach resonance \cite{FR}. In the following we will show that the non-Hermitian model can be solved exactly both in periodic boundary condition (PBC) and OBC for uniform potential $\phi(x)=\phi$. 

We start from the Schr\"odinger equation, $\hat{H}\Psi\left(x_{1}, \ldots, x_{N}\right)=E\Psi\left(x_{1}, \ldots, x_{N}\right)$ and then we write the  many-body wavefunction $\Psi\left(x_{1}, \ldots, x_{N}\right)$ in the following form 
\begin{eqnarray} 
\Psi\left(x_{1}, \ldots, x_{N}\right)&=& \sum_{{\bf P}} \psi\left(x_{p_{1}}, x_{p_{2}}, \ldots, x_{p_{N}}\right) \nonumber \\
& \times& \Theta\left(x_{p_{1}}\leq x_{p_{2}} \cdots\leq x_{p_{N}}\right)
\end{eqnarray} 
where $p_{1}, p_{2}, \ldots, p_{N}$ presents the one of permutations of the set $1, \ldots, N$, and $\sum_{\bf P}$ is the summation of all permutations. $\Theta\left(x_{p_{1}}\leq \cdots\leq x_{p_{N}}\right)=\theta\left(x_{p_{N}}-x_{p_{N-1}}\right) \cdots \theta\left(x_{p_{2}}-x_{p_{1}}\right)$ where $\theta(x-y)$ is the step function. Since the wavefunction is symmetric under the interchange of coordinates, one just needs to calculate $\psi\left(x_{1},x_{{2}}, \ldots, x_{N}\right)$ in any region $x_{{1}}\leq x_{{2}}\leq  \cdots\leq x_{{N}}$, which satisfies
\begin{eqnarray} 
\left(\sum_{j=1}^{N} \Big[-i\frac{\partial}{\partial x_{j}}+i\phi\Big]^2+2 c \sum_{i<j} \delta\left(x_{i}-x_{j}\right)\right) \psi\left(x_{1}, \ldots, x_{N}\right)\nonumber \\
= E \psi\left(x_{1}, \ldots, x_{N}\right).~~~~~
\end{eqnarray} 
The $\delta$-function interaction gives rise to the contact condition
\begin{eqnarray}
\left(\frac{\partial}{\partial x_{j+1}}-\frac{\partial}{\partial x_{j}}\right) \psi\left(\cdots x_{j}, x_{j+1} \cdots\right)\big|_{x_{j+1}=x_{j}}\nonumber \\
=c \psi\left(\cdots x_{j}, x_{j+1} \cdots\right)\big|_{x_{j+1}=x_{j}}. \label{contact} 
\end{eqnarray}

We first consider the PBC, i.e.,
\begin{eqnarray} 
\psi\left(x, x_{2}, \ldots, x_{N}\right)=\psi\left(x_{1}, x_{2}, \ldots, x+L\right), \label{PBC}
\end{eqnarray} 
with system size $L$.
According to the Bethe ansatz solutions, the wavefunction takes linear superposition of plane waves 
\begin{eqnarray} 
\psi\left(x_{1}, x_{2}, \ldots, x_{N}\right)=\sum_{{\bf P}}A_{{\bf P}} \exp \Big(\sum_{j=1}^{N} ik_{p_{j}} x_{j}\Big), \label{PBC_wavefuntion}
\end{eqnarray} 
where $k_j$'s denote the quasimomentum of bosons. Combining the wavefunction (\ref{PBC_wavefuntion}) into contact condition (\ref{contact}) and boundary condition (\ref{PBC}), quasimomenta satisfy the Bethe ansatz equations (BAEs)
\begin{eqnarray} 
\exp \left(i k_{j} L\right)=\prod_{l=1(\neq j)}^{N}  \frac{i k_{j}-i k_{l}-c}{i k_{j}-i k_{l}+c}, 
\end{eqnarray} 
and the eigenenergy is given by $E=\sum_{j=1}^{N} (k_{j}+i\phi)^{2}$.
The solution of quasimomenta and the corresponding eigenfunction are independent of $\phi$, but the spectrum is complex, which is similar to the single-particle model in PBC as discussed in appendix \ref{App1}.

We then turn to the OBC 
\begin{eqnarray} 
\psi\left(0, x_{2}, \ldots, x_{N}\right)=\psi\left(x_{1}, x_{2}, \ldots, L\right)=0. \label{OBC} 
\end{eqnarray}
It had been shown that the Lieb-Liniger model ($\phi=0$) in OBC can also be exactly solved \cite{LL_OBC1,LL_OBC2,LL_OBC3,LL_OBC4}. We will show that the non-Hermitian case can be solved exactly. To solve the non-Hermitian model in OBC, we require the non-Bloch wavefunction obtained from the single-particle model in OBC. And then we can construct the many-body wavefunction by means of Bethe ansatz form
\begin{align} 
\psi\left(x_{1}, x_{2}, \ldots, x_{N}\right)=\sum_{{\bf P}, r_{1}, \ldots, r_{N}} A_{\bf P} \exp \Big(\sum_{j=1}^{N} (i r_{j} k_{p_{j}} x_{j}+\phi x_j)\Big), \label{OBC_wavefuntion} 
\end{align} 
where $r_{j}=1$ ($r_{j}=-1$) indicates the plane wave of the $j$th boson moving toward right or left. We emphasis that Bethe ansatz wavefunction (\ref{OBC_wavefuntion}) is the superposition of non-Bloch wavefunctions instead of plane waves. The traditional ansatz for Hermitian systems in OBC, i.e. $\psi\left(x_{1}, x_{2}, \ldots, x_{N}\right)=\sum_{{\bf P}, r_{1}, \ldots, r_{N}} A_{\bf P} \exp\big(\sum_{j=1}^{N} i r_{j} k_{p_{j}} x_{j}\big)$ can not solve the non-Hermitian Hamiltonian (\ref{H}).
Based on the wavefunction (\ref{OBC_wavefuntion}) together Eq. (\ref{OBC}) and Eq. (\ref{contact}), one can derive the BAEs in OBC
\begin{eqnarray} 
\exp \left(i 2 k_{j} L\right)=\prod_{l=1(\neq j)}^{N} \frac{i k_{j}-i k_{l}-c}{i k_{j}-i k_{l}+c}\frac{i k_{j}+i k_{l}-c}{i k_{j}+i k_{l}+c},\label{BAEs_OBC}
\end{eqnarray} 
and the energy eigenvalue is given by $E=\sum_{j=1}^{N} k_{j}^{2}$. Physically, the second fraction on the right-hand side of Eq. (\ref{BAEs_OBC}) comes from the reflection at boundaries. Distinct from the situation in PBC, the spectrum in OBC is independent of $\phi$ and takes real, but the wavefunction depends on $\phi$. As a consequence, the many-body eigenfunction in OBC exhibits NHSE in entire area of interaction $c\in(-\infty,\infty)$. The following sections will elaborate on the properties of NHSE in distinct interaction regions, including repulsive and attractive interactions.
 
\section{NHSE in repulsive interaction}
\label{Sec. repulsive}
In this section, we study the NHSE in repulsive interaction. When $c>0$, the system is in a scattering state, and the solution of quasimomenta is real and unique. To analyze the solutions of BAEs, we first take the logarithm of Eq. (\ref{BAEs_OBC}) that leads to
\begin{align} 
k_{j} L= \pi{I}_{j}+\sum_{l=1(\neq j)}^{N}\left(\arctan \frac{c}{k_{j}-k_{l}}+\arctan \frac{c}{k_{j}+k_{l}}\right), \label{BAEs_log}
\end{align} 
where $\left\{I_{j}\right\}$ represents quantum number that takes a set of integers. For the ground state, we have $I_{j}=1~(1 \leq j \leq N)$ which can be found in non-interacting limit where all bosons condense on $k_j=\pi/L$. 
For a fixed $\left\{n_{j}\right\}$, one can solve Eq. (\ref{BAEs_log}) to determine quasimomenta for arbitrary interaction strength.
 
After some calculations, we can derive the explicit expression of the eigenfunction as 
\begin{align} 
\psi\left(x_{1}, x_{2}, \ldots, x_{N}\right)&= \sum_{\bf P} A_{\bf P} \epsilon_{\bf p} \exp \left[i\left(\sum_{l<j}^{N-1} \Omega_{p_{j} p_{l}}\right)\right]\nonumber \\
&\times\sin \left(k_{p_{1}} x_{1}\right)\exp \left(i k_{p_{N}} L\right)\prod_{j=1}^N\exp \left(\phi x_j\right)\nonumber \\ 
\times \prod_{1<j<N} \sin &\Big(k_{p_{j}} x_{j}-\sum_{l<j} \Omega_{p_{l} p_{j}}\Big) \sin \Big[k_{p_{N}}\left(L-x_{N}\right)\Big], 
\end{align} 
where $A_{\bf P}=\prod_{j<l}^{N}\left(i k_{p_{j}}-i k_{p_{l}}-c\right)\left(i k_{p_{j}}+i k_{p_{l}}-c\right)$ and $\Omega_{j l}=\arctan \frac{c}{k_{j}+k_{l}}-\arctan \frac{c}{k_{j}-k_{l}}$. Here the sign factor $\epsilon_{\bf p}$ takes $+1$ $(-1)$ relying on even (odd) permutations of ($p_{1}, p_{2}, \ldots, p_{N}$). Apparently, the wavefunction meets boundary conditions (\ref{OBC}). Thus, the eigen-energy and wavefunction can be determined immediately once the BAEs are solved. 

\begin{figure}[htp]
	\centering
	\includegraphics[width=8.5cm]{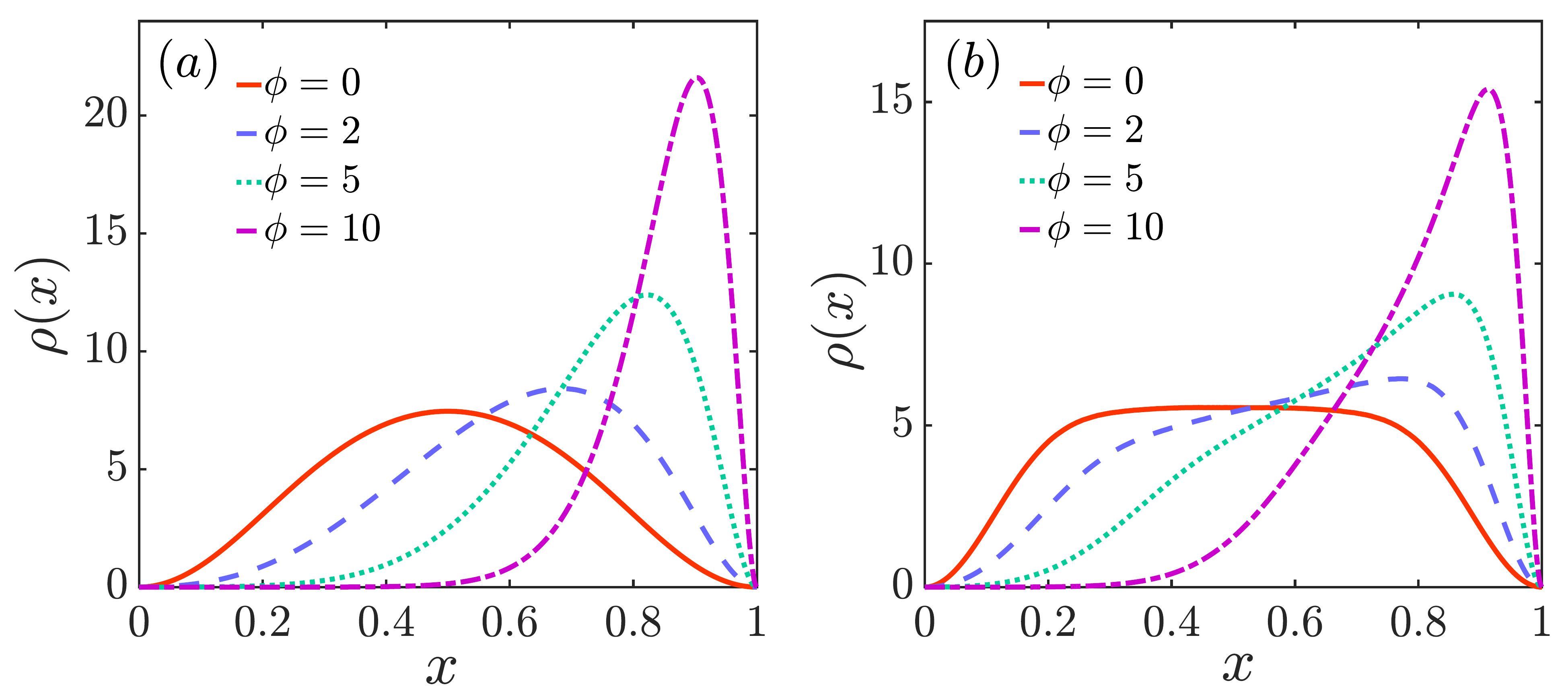}
	\includegraphics[width=8.5cm]{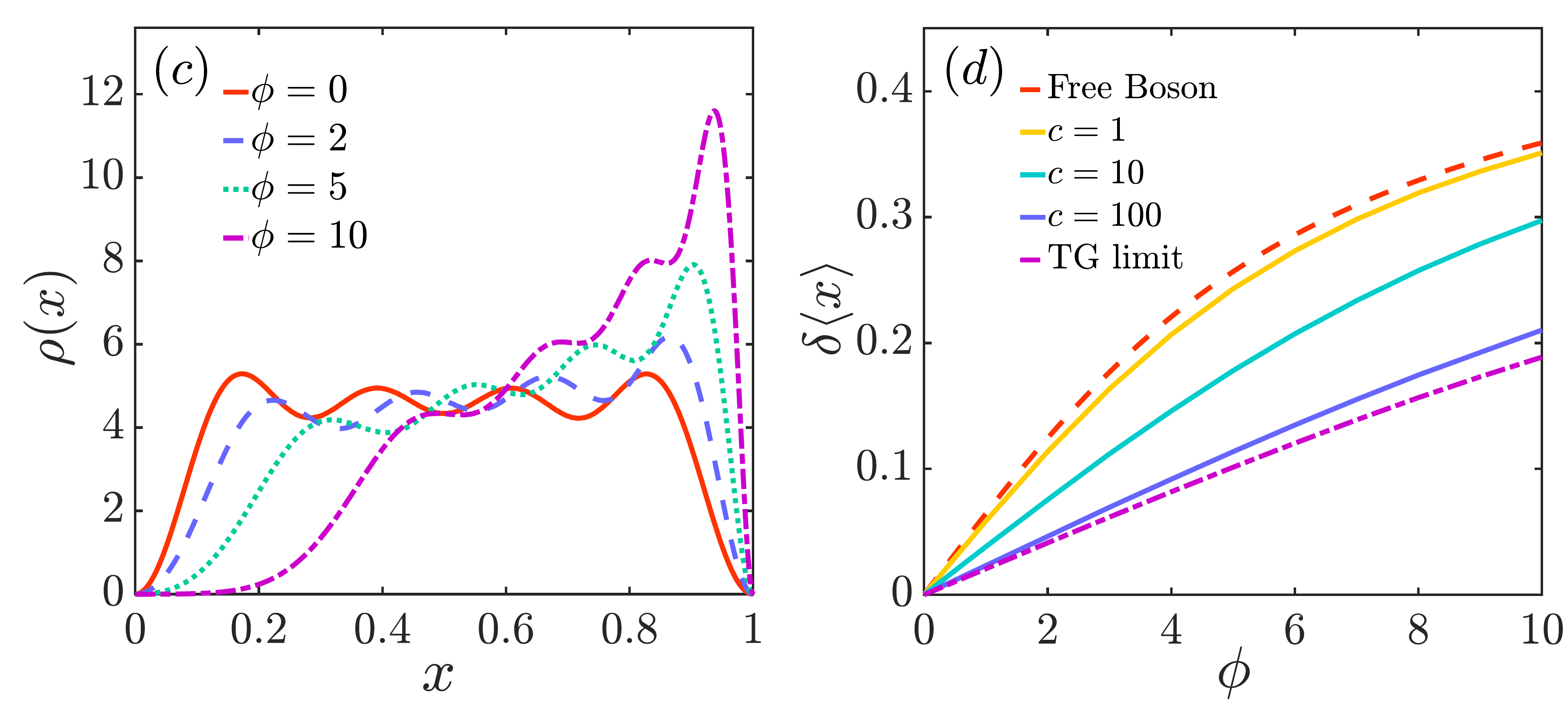}
	\caption{NHSE in repulsive interacting Bosons. Groundstate density distributions $\rho(x)$ in real space for different $\phi$ at $c=1$ (a), $c=10$ (b) and $c=100$ (c), respectively. (d) The deviation of mean position $\delta\langle x\rangle$ as a function of $\phi$ for different interaction strengths from non-interacting (dashed line) to TG limit (dot-dashed line). Here we choose $L=1$ as length unit and set $N=4$.}
	\label{Fig1}
\end{figure}

To investigate the NHSE, we calculate the density distribution in real space 
\begin{align}
\rho(x)=\frac{N \int_{0}^{L} d x_{2} \cdots d x_{N}\left|\Psi\left(x, x_{2},  \ldots, x_{N}\right)\right|^{2}}{\int_{0}^{L} d x_{1} \cdots d x_{N}\left|\Psi\left(x_{1}, x_{2}, \ldots, x_{N}\right)\right|^{2}}.
\end{align}
In Fig. \ref{Fig1}, we plot density distributions for distinct potential $\phi$ at weak ($c=1$), mediate ($c=10$) and strong interaction ($c=100$).  
Notably, the density distribution gradually tends to the right boundary with the increasing of $\phi$. Meanwhile, the repulsive interaction widens the density profile. In order to characterize the degree of NHSE qualitatively, we define the deviation of mean position similar to the charge distribution    
\begin{align} 
\delta\langle x\rangle=\frac{1}{N}\int_{0}^{L}x\Big[\rho(x)-\rho_0(x)\Big]dx,
\end{align} 
where $\rho_0(x)$ denotes the density distribution at $\phi=0$. 
The larger $\delta\langle x\rangle$ is, the stronger degree of the NHSE will be.
One can see from Fig. \ref{Fig1}(d) that the NHSE is suppressed as the interaction strength grows. It is physically reasonable that the repulsive interaction prevents bosons from clumping together, thereby effectively suppressing NHSE.
However, the NHSE always exists even in the Tonks-Giradeau (TG) limit $c=\infty$. In TG limit, the ground state solution of the BAEs (\ref{BAEs_log}) is $k_j=j\pi$ ($j=1,2,\cdots, N$) that the bosons look like free spinless fermions, and the corresponding wavefunction takes the form of
\begin{align}
\Psi\left(x_{1}, x_{2}, \ldots, x_{N}\right)=& \sum_{\bf P} \theta\left(x_{p_{N}}-x_{p_{N-1}}\right) \cdots \theta\left(x_{p_{2}}-x_{p_{1}}\right) \nonumber \\
\times	& \prod_{j=1}^{N} \sin \left(j\pi x_{p_{j}}/L\right) \exp \left(\phi x_{p_{j}}\right),
\end{align}
which leads to finite $\delta\langle x\rangle$ as shown in Fig. \ref{Fig1}(d). 

\begin{figure}[htp]
\centering
\includegraphics[width=8.5cm]{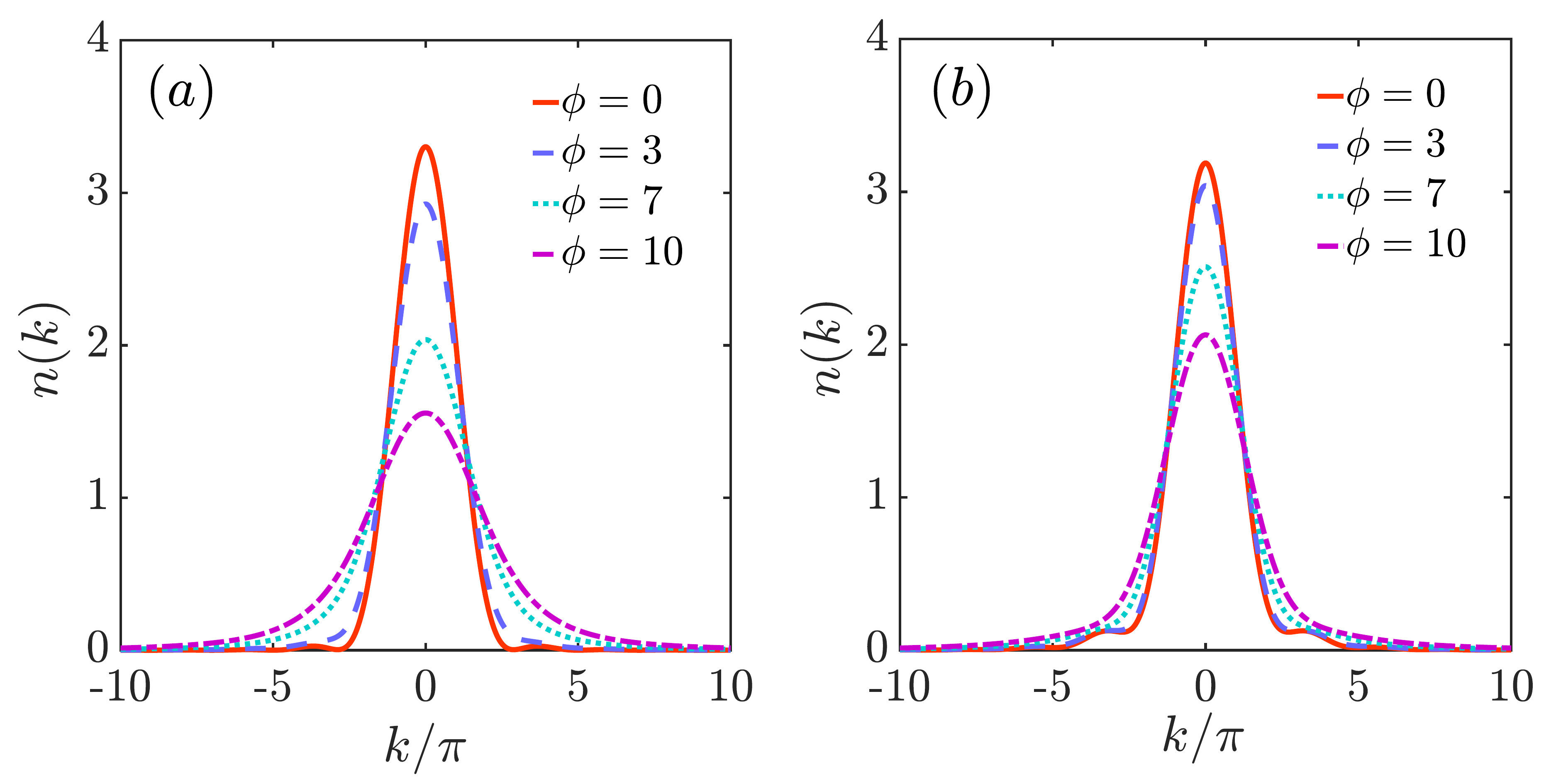}
\includegraphics[width=8.5cm]{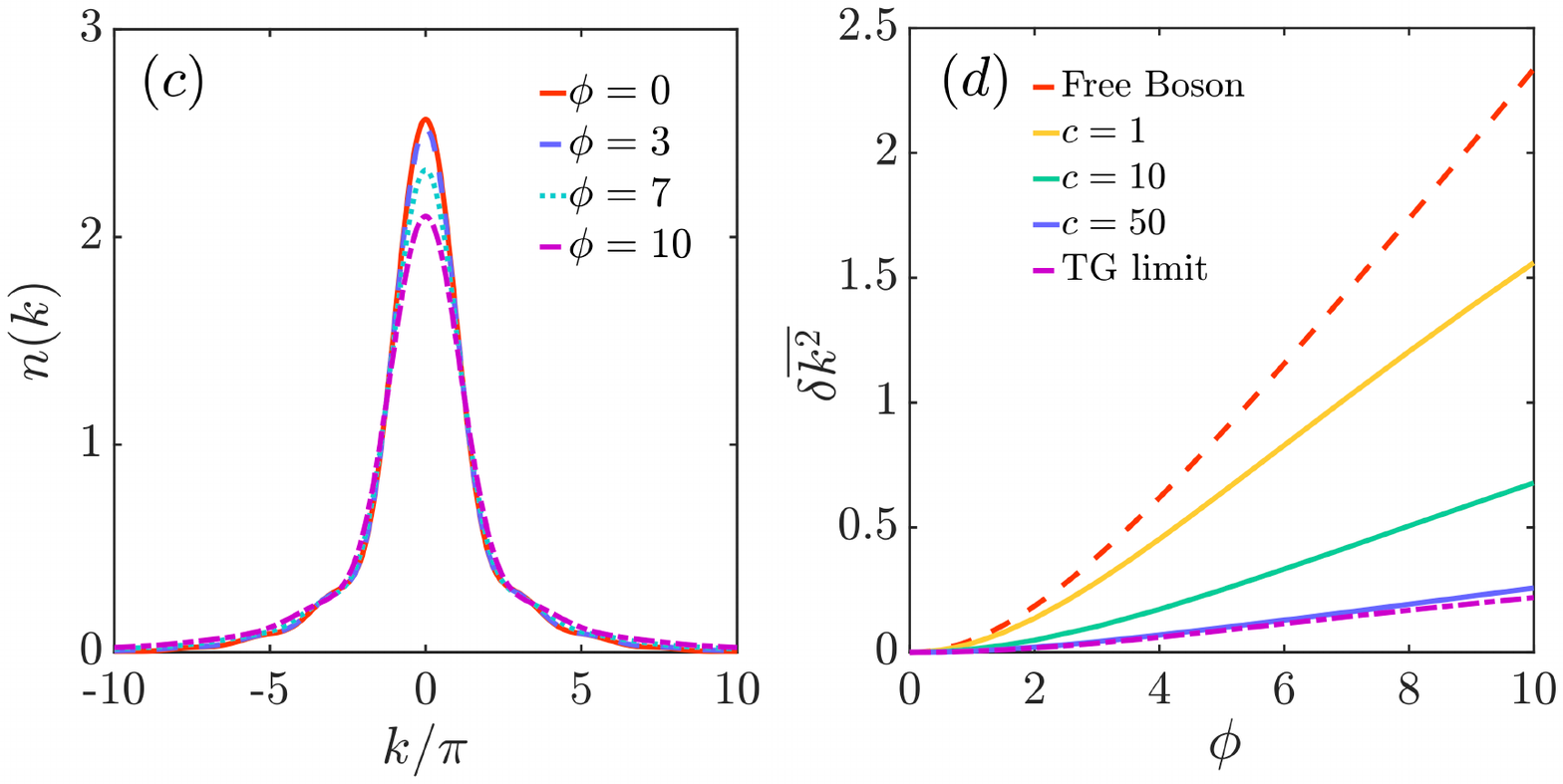}
\caption{Groundstate momentum distributions $n(k)$ for different $\phi$ at $c=1$ (a), $c=10$ (b) and $c=50$ (c), respectively. (d) The deviation of momentum distribution $\delta\langle x\rangle$ as a function of $\phi$ for different interaction strengths from non-interacting (dashed line) to TG limit (dot-dashed line). Here we choose $L=1$ as length unit and set $N=4$. }
\label{Fig2}
\end{figure}

We can also characterize the magnitude of NHSE via momentum distribution  $n(k)=\frac{1}{2 \pi} \int_{0}^{L}dx\int_{0}^{L}dx^{\prime}\varrho\left(x, x^{\prime}\right) e^{-i k\left(x-x^{\prime}\right)}$ where the single-particle density matrix $\varrho\left(x, x^{\prime}\right)$ is defined as
\begin{equation}
\varrho\left(x, x^{\prime}\right)=
\frac{N \int_{0}^{L} d x_{2} \cdots d x_{N} \Psi^{*}\left(x, x_{2}, \ldots, x_{N}\right) \Psi\left(x^{\prime}, x_{2}, \ldots, x_{N}\right)}{\int_{0}^{L} d x_{1} d x_{2}\cdots d x_{N}\left|\Psi\left(x_{1}, x_{2}, \ldots, x_{N}\right)\right|^{2}}.
\end{equation}
The momentum distribution width is enhanced as $\phi$ increases, as shown in Fig. \ref{Fig2} (a), Fig. \ref{Fig2} (b) and Fig. \ref{Fig2} (c). The broadening momentum distribution corresponds to the NHSE in real space, where particles concentrate on the boundary with a narrower density profile according to the Fourier transformation. Specifically, Fig. \ref{Fig2}(d) plots the deviation of momentum distribution width $\delta\overline{k}^2=\overline{k^2}-\overline{k^2}_{\phi=0}$ with 
$\overline{k^2}=\int_{-\infty}^{\infty}k^2 n(k)dk$ as a function of $\phi$ at different interaction strengths where the suppression of the momentum distribution width growth by repulsive interaction can be visualized.  

\section{NHSE in attractive interaction}
\label{Sec. attractive}
Sec. \ref{Sec. repulsive} deals with the situation of repulsive interaction. We now turn to study the NHSE in attractive interaction. What should be pointed out is that unlike the case of repulsive interaction, the solution of BAEs in attractive interaction is not unique, which contains bound state and scattering states. Physically, there exists a bound state when bosons attract each other, reflected in the complex solution of quasimomenta $\{k_j\}$ in BAEs. In weak interaction region, the ground state solution of quasimomenta consists of $N/2$ ($N$ is even number) pairs of conjugate complex roots, i.e., $N/2$ dimers $k_{2j-1}=\alpha_j-i\beta_j$, $k_{2j}=\alpha_j+i\beta_j$ ($j=1,\cdots,N/2$).   
The state made up of $N/2$ dimers is also called $N/2$ two-string. The $M$-string state is defined by $M$ quasimomenta sharing the same real part but unequal conjugate imaginary parts, which can be labeled as $k_j=\alpha+\beta_j$ ($j=1,\cdots,M$).
According to the bound state solution of BAEs, as the attractive interaction strength increases, the ground state evolves from a $N/2$ two-string state gradually to an intermediate state characterized by an $M$-string ($N>M>2$) plus $(N-M)/2$ two-string and finally turns to $N$-string state. 

\begin{figure}[!h]
	\centering
	\includegraphics[width=8.5cm]{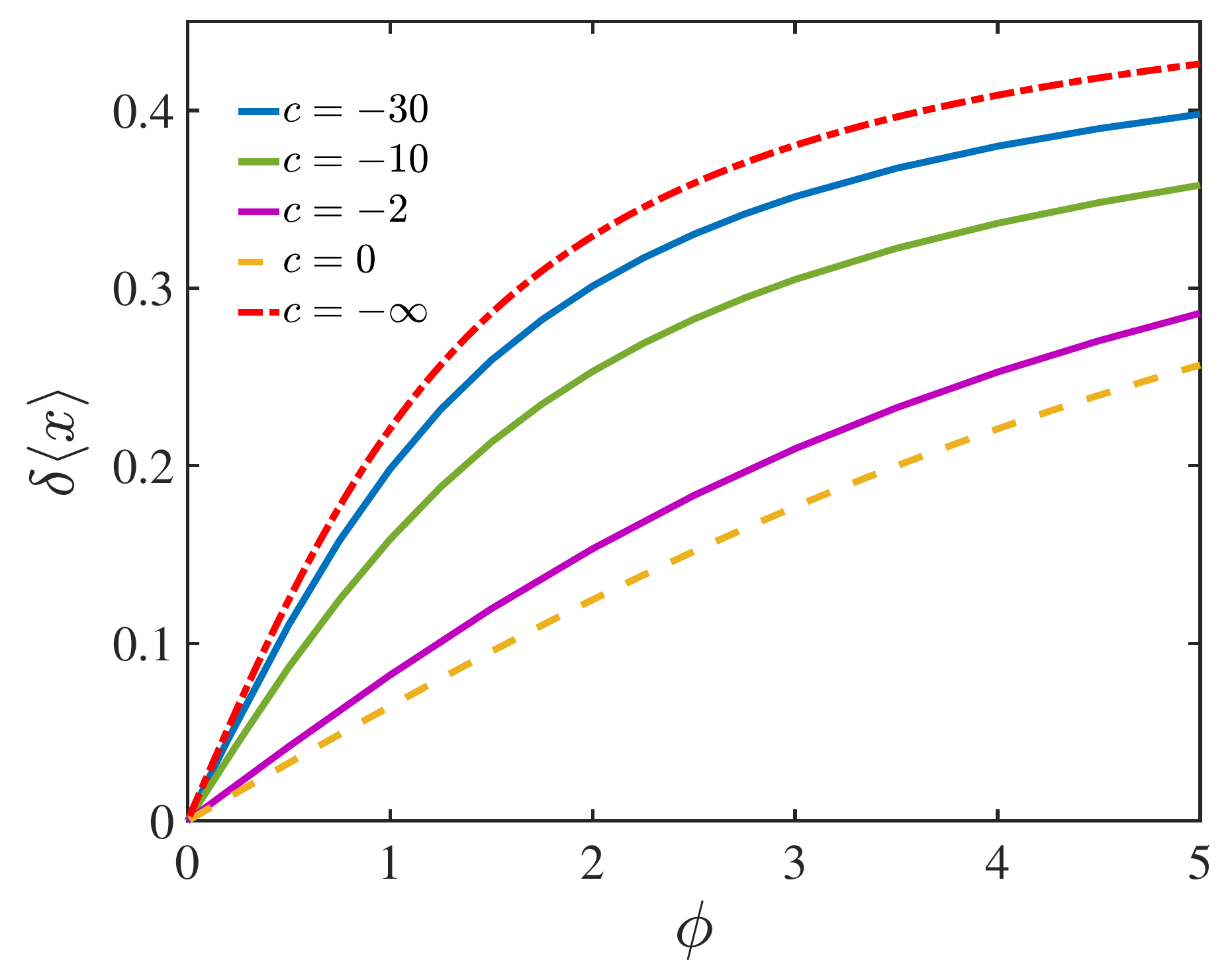}
	\caption{NHSE in attractive interacting Bosons for bound state. The deviation of mean position $\delta\langle x\rangle$ as a function of $\phi$ for different interaction strengths from non-interacting (dashed line) to infinite attractive limit (dot-dashed line). Here we choose $L=1$ as length unit and set $N=4$.}
	\label{Fig3}
\end{figure}

The bound state solution of BAEs shows that the NHSE exists in the entire range of interaction strength $c\leq0$ as shown in Fig. \ref{Fig3}. Moreover, the deviation $\delta\langle x\rangle$ tends to increase with the growth of attractive strength. This is because the bosons tend to clump together with increasing attraction among them. We can also understand the NHSE in attractive interaction in terms of quantitative analysis in two limits. For non-interacting limit, all bosons occupy the single-particle ground state $k_j=\pi/L$ and the corresponding groundstate wave function is simply written as $\Psi\left(x_{1}, x_{2}, \ldots, x_{N}\right)=\prod_{j=1}^{N} \sin \left(j\pi x_{j}/L\right) \exp \left(\phi x_{j}\right)$ which produces the deviation of mean position $\delta\langle x\rangle=\frac{L}{2}\left[\coth(\phi L)-\frac{1}{\phi L}-\frac{2 \phi L}{\pi^{2}+\phi^{2}L^2}\right]$. In strongly attractive limit $c=-\infty$, the system forms a $N$-body bound state corresponding to the $N$-string state. At this point, the quasimomentum distribution is given by $k_{j}=K/N+i(N+1-2 j)c/2, (j=1,2,\cdots N)$ where $K$ signifies the total momentum, which is determined by the following transcendental equation 
\begin{equation}
KL= \pi I+2\sum_{j=1}^{N-1}\arctan \frac{j cN }{2 K}
\end{equation}
with $I=N$ for the ground state. We can see that in the limit of $c \rightarrow-\infty$, the total momentum is $K=\pi/L$ which indicates each boson has the real part $\pi/NL$ in the strong attractive limit and the corresponding ground state wavefunction to be written as 
\begin{equation}
\Psi\left(x_{1}, x_{2}, \ldots, x_{N}\right)=\sin \left(\pi x/L\right) \exp \left(N\phi x\right), \label{N_bound}
\end{equation}
where $x=\frac{x_1+x_2+\cdots+x_N}{N}$ denotes the center-of-mass coordinate. The wave function (\ref{N_bound}) gives a clear physical picture that the system forms a giant molecule of $N$ bosons involving the movement of center-of-mass. And similarly, the deviation of mean position derived from (\ref{N_bound}) yields 
\begin{align}
\delta\langle x\rangle=\frac{L}{2}\left[\coth\Phi L-\frac{1}{\Phi L}-\frac{2 \Phi L}{\pi^{2}+\Phi^{2}L^2}\right], 
\end{align}
with $\Phi=N\phi$. We can see that the $\delta\langle x\rangle$ of infinitely attractive bosons is similar to the one of free bosons in form but contains a factor $N$. Physically, it comes from the Bose enhancement that all bosons tend to locate at the same position.

\begin{figure}[h]
	\centering
	\includegraphics[width=8.5cm]{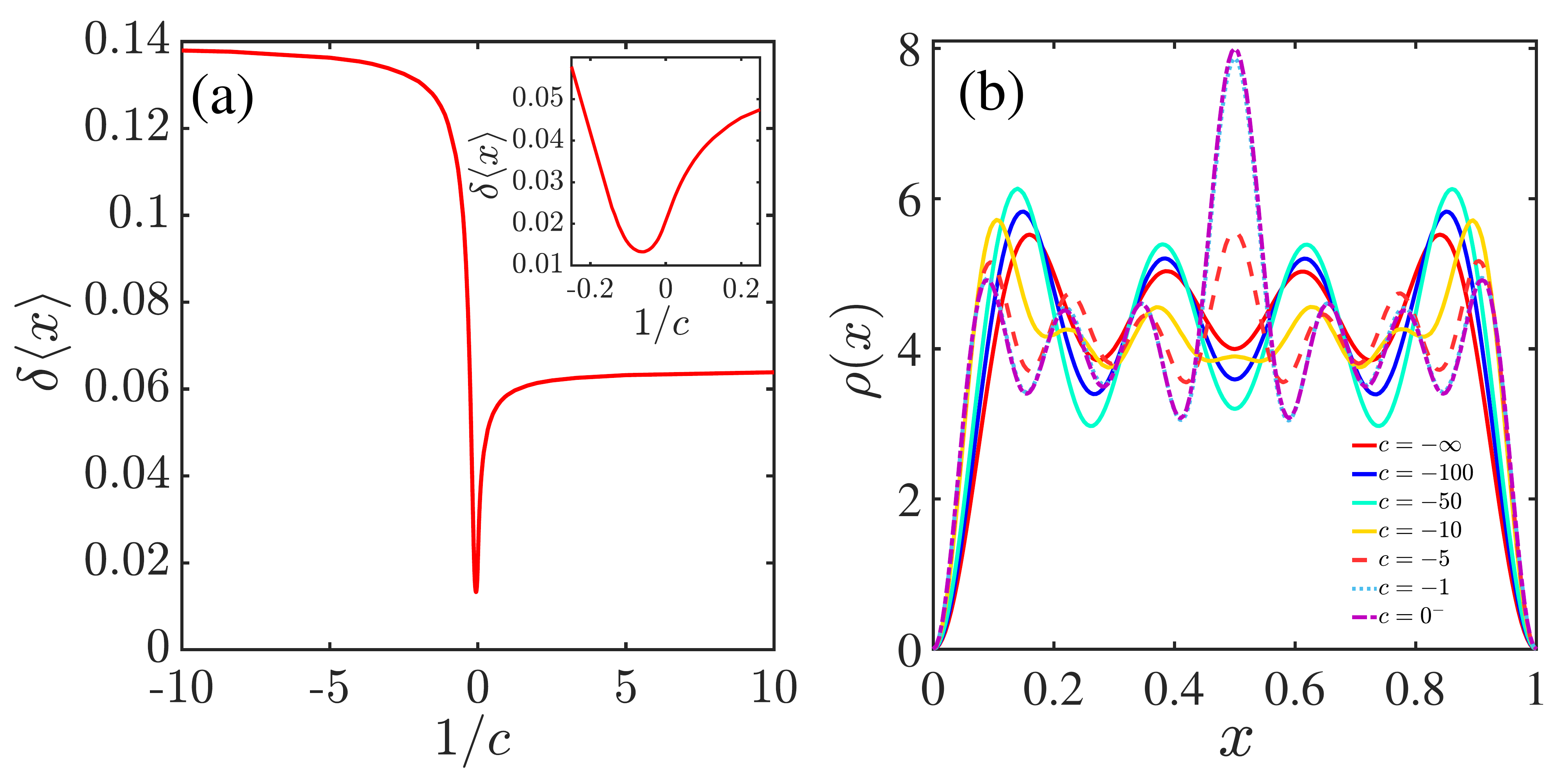}
	\caption{Panel (a): The deviation of mean position $\delta\langle x\rangle$ as a function of inverse interaction strength at $\phi=1$ for scattering state.  Panel (b): Density distributions of the scattering state for different attractive interaction strengths from non-interacting (dashed line) to infinite attractive limit (dot-dashed line). . 
	Here we choose $L=1$ as length unit and set $N=4$. }\label{Fig4}
\end{figure}

Next, we turn to address the case of scattering states on the attractive side. In fact, there are real solutions corresponding to scattering states in BAEs 
(\ref{BAEs_OBC}) for $c<0$ although its ground state is a string solution (bound state). In the strong interaction regime, this kind of scattering state is referred to super-Tonks-Girardeau (STG) gas in the literature \cite{STG1,STG2,STG3,STG4,STG5}.
For the purpose of analysis, we alternatively rewrite the BAEs (\ref{BAEs_OBC}) as the following logarithm form 
\begin{align} 
k_{j} L= \pi I_{j}-\sum_{l=1(\neq j)}^{N}\left(\arctan \frac{k_{j}-k_{l}}{|c|}+\arctan \frac{k_{j}+k_{l}}{|c|}\right). \label{BAEs_log2}
\end{align}  
Here we choose $I_{j}=j$, such that the quasimomentum distribution
in $c\rightarrow-\infty$ determined from Eq. (\ref{BAEs_log2}) connects to the TG limit from Eq. (\ref{BAEs_log}) where $k_j=j\pi/L$ ($j=1,2,\dots,N$). Under this circumstance, one would obtain the deviation of mean position $\delta\langle x\rangle$ as a continuous function of $1/c$ as shown in Fig. \ref{Fig4}(a). We can recognize that in contrast to the bound state on the attractive side, the scattering state exhibits a non-monotonic behavior where $\delta\langle x\rangle$ decreases first and then increases as the attractive strength weakens. And also the transition point (the inset of Fig. \ref{Fig4} (a)) locates at the STG regime ($-1\ll1/c<0$). 
The non-monotonic behavior is closely related to the density distribution on the attractive side.
In the limit of $c\rightarrow-\infty$, the density profile is identical to the one in the TG limit displaying $N$ peaks with near equidistant. Then, the density distribution gradually migrates toward boundaries as the attraction decreases, suppressing the NHSE. However, as the attraction is getting weaker, the density tends to concentrate in the center of potential with a larger weight which can magnify the NHSE. Finally, in weak attraction limit, there emerge $2N-1$ peaks in the density distribution as shown in Fig. \ref{Fig4}(b). This is because the bosons occupy $N$ lowest odd single-particle orbitals $k_j=(2j+1)\pi/L$ when $c\rightarrow0^-$.

\section{Summary and Outlook}
\label{Summary}

In summary, through the example of non-Hermitian 1D interacting bose gas, we investigated the NHSE in a non-Hermitian many-body system. Utilizing Bethe ansatz, we obtained the exact solutions, including quasimomenta, eigen-energies, and wavefunctions of the system in OBC. The NHSE can be characterized by density distribution and momentum distribution. Our calculations show that the NHSE exists in the entire interaction regime and displays distinct responses to the interaction effect.
The main conclusions are summarized as follows:

(I) For repulsive interaction $c>0$, the NHSE is gradually suppressed as interaction strength increases but does not vanish even in the TG limit ($c\rightarrow+\infty$). 

(II) For the bound state in attractive interaction $c<0$, the system consists of $N$ bosons from the 2-string state into the $N$-string state as the attractive interaction strength grows. And in this process, the NHSE is enhanced.   

(III) For scattering state in attractive interaction $c<0$, we find a non-monotonic behavior in the deviation of mean position $\delta \langle x\rangle$ where the NHSE is first suppressed and then enhanced during interaction weakens. 

Our non-Hermitian many-body model can, in principle, be realized in current cold atom experiments, which offer an ideal platform to observe the NHSE. Thanks to the flexible tunability, cold atomic gases have realized tunable non-reciprocal model (with unequal hopping strength) through a dissipative Aharonov-Bohm ring \cite{BoYan_AB_ring} and observed dynamic signatures of the NHSE \cite{Skin_Exp9}. Our non-Hermitian Lieb-Liniger model describes the low-filling regime of the 1D Bose-Hubbard mode, which could be realized from the aforementioned non-reciprocal model by adding tunable interactions. The NSHE in continuum space is proposed to be observed via dynamic measurements \cite{Skin_Theory19}. In situations with and without NSHE respectively, a right-moving wave package will either be localized near the right boundary after touching it, or reflected as in the usual cases. We believe our model can also be realized in a similar way by adding proper interactions.

This paper provides some preliminary exploration of the NHSE in many-body systems from the view of exactly solvable many-body models. Our results build on the rapidly expanding field of non-Hermitian physics. Considerably more work will be desired to explore the interplay between the NHSE and other kinds of exactly solvable many-body systems and further accomplish more comprehensive investigations for future research.

\section*{Acknowledgements}
We would like to thank Pengfei Zhang and Tianshu Deng for helpful discussions. We also thank Hui Zhai for his valuable comment. This work is supported by Beijing Outstanding Young Scientist Program held by Hui Zhai. YH was supported by NSF of China under
Grants No. 11774026. LP acknowledges support from the project funded by the China Postdoctoral Science Foundation (Grant No. 2020M680496). 
\appendix

\section{Single-particle non-Hermitian skin effect in tight-binding model}\label{App1}
In this appendix we give a brief introduction to the non-Hermitian skin effect at a single-particle level. We consider a one-dimensional tight-binding model with unequal hopping, whose Hamiltonian is written as  
 
\begin{eqnarray} 
	\hat{H}=t_{1} \sum_{n}|n+1\rangle\langle n |+t_{2} \sum_{n}| n\rangle\langle n+1|, \label{HN_model}
\end{eqnarray}
where $t_{1}>t_{2}$ denote the unequal hopping amplitudes. This non-Hermitian model with disorder potential was first introduced by Hatano and Nelson \cite{HN_Model} to investigate localization transition and is investigated in ultracold atoms \cite{XuYong,RenZhang}. Now we solve its eigenvalues and eigenvectors. Under periodic boundary condition (PBC), the momentum is a good quantum number ($k=$ $0,2 \pi/L, \ldots, 2 \pi(L-1)/L$). Then the eigenenergies and eigenstates are given by  
\begin{eqnarray} 
E_{k}=t_{1} e^{-i k}+t_{2} e^{i k}, \quad\left|E_{k}\right\rangle=\frac{1}{\sqrt{L}} \sum_{n} e^{i k n}|n\rangle,
\end{eqnarray}
from which we can see the eigenvector is nothing but the Bloch state which is identical to Hermitian case but the spectrum becomes complex.
%

Under open boundary condition (OBC), the momentum is not good quantum number anymore. Expanding the eigenvector in real space $|E\rangle=\sum_{n} \psi_{n}|n\rangle$, we have
\begin{eqnarray} 
E \psi_{n}=t_{1} \psi_{n-1}+t_{2} \psi_{n+1}, \label{Ener_HNmodel}
\end{eqnarray}
with OBC $\quad \psi_{0}=\psi_{L+1}=0$. We introduce a non-Bloch wavefunction $\psi_{n}=A_{1} \beta_{1}^{n}+A_{2} \beta_{2}^{n}$, and then the eigenvalues are determined by 
\begin{eqnarray} 
E(\beta)=t_{1} / \beta_{a}+t_{2} \beta_{a}, 
\end{eqnarray} 
where $a=1,2$. The boundary condition $\quad \psi_{0}=0,~\psi_{L+1}=0$ give 
\begin{eqnarray} 
A_{1}+A_{2}=0, \quad A_{1} \beta_{1}^{L+1}+A_{2} \beta_{2}^{L+1}=0,
\end{eqnarray}
which will lead to the condition $\beta_{1}^{L+1}=\beta_{2}^{L+1}$. Together with $\beta_{1} \beta_{2}=t_{1}/t_{2}$, we can obtain
\begin{eqnarray} 
\beta_{1}=\sqrt{t_{1} / t_{2}} e^{i \theta_{m}}=\beta_{2}^{*}, \quad \theta_{m}=\pi m /(L+1),
\end{eqnarray} 
with $m=1,2, \ldots, L$. Therefore, we can derive the eigenenergies in OBC
\begin{eqnarray} 
E_{m}=2 \sqrt{t_1 t_2} \cos \theta_{m},
\end{eqnarray} 
and associated eigenstates 
\begin{eqnarray} 
\left|E_{m}\right\rangle=\sum_{n} e^{n\phi}\sin n\theta_{m}|n\rangle, \label{Wavefunction_OBC}
\end{eqnarray} 
where $\phi=\ln\sqrt{t_{1}/t_{2}}$. Note that the spectrum in OBC is always real. Meanwhile, the non-Bloch wavefunction consists of the standing wave $\sin (n\theta_{m})$ with an amplification factor $e^{n\phi}$ exhibiting exponential accumulation at the boundary. This feature is called the non-Hermitian skin effect. The non-Bloch wavefunction in the eigenstate \eqref{Wavefunction_OBC} will be the starting point for solving non-Hermitian many-body systems. 
 


\end{document}